\documentclass[aps,amsmath,showpacs,amsfonts,10pt]{revtex4}
\usepackage{epsfig,graphicx}
\usepackage[english]{babel}
\usepackage{amsfonts}
\usepackage{amsmath}
\usepackage{latexsym}
\usepackage{graphics,bm}
\usepackage{natbib}
\usepackage{dcolumn}
\usepackage{bm}
\usepackage{rotating}

\begin{document}

\title{Dynamical Casimir Effect in Superradiant Light Scattering by Bose-Einstein Condensate in an Optomechanical Cavity}

\author{ Sonam Mahajan$^{1}$ , Neha Aggarwal$^{1,2}$, Aranya B Bhattacherjee$^{2,3}$ and ManMohan$^{1}$}

\address{$^{1}$Department of Physics and Astrophysics, University of Delhi, Delhi-110007, India} \address{$^{2}$Department of Physics, ARSD College, University of Delhi (South Campus), New Delhi-110021, India} \address{$^{3}$School of Physical Sciences, Jawaharlal Nehru University, New Delhi-110067, India}

\begin{abstract}
We investigate the effects of dynamical Casimir effect in superradiant light scattering by Bose-Einstein condensate in an optomechanical cavity. The system is studied using both classical and quantized mirror motions. The cavity frequency is harmonically modulated in time for both the cases. The main quantity of interest is the number of intracavity scattered photons. The system has been investigated under the weak and strong modulation. It has been observed that the amplitude of the scattered photons is more for the classical mirror motion than the quantized mirror motion. Also, initially, the amplitude of scattered photons is high for lower modulation amplitude than higher modulation amplitude. We also found that the behaviour of the plots are similar under strong and weak modulation for the quantized mirror motion.
\end{abstract}


\maketitle

\section{Introduction}

The long-range coherence of Bose-Einstein condensate (BEC) offers the possibility to study the collective motion induced by external radiation beams \citep{bos}. In specific, it has been observed that the coherent center-of-mass motion of atoms in a condensate driven by a highly detuned laser gives rise to the superradiant Rayleigh scattering and matter-wave amplification \citep{ino, sch1, koz}. The matter-wave grating produced in these experiments is due to the coherent superposition of atomic momentum states which is identical to the one formed in Bragg scattering experiments in which the condensate atoms are diffracted by a standing wave of light \citep{ste, koz1}. The collective atomic recoil lasing (CARL) \citep{bon, bon1} process shows the collective instability which causes the spontaneous formation of a regular density grating in a BEC which was first demonstrated in superradiant Rayleigh scattering experiments \citep{ino} and matter-wave amplification \citep{koz}. The coherence in the atomic momentum states plays an important role in all these experiments. The CARL process can be supressed by the effects such as spontaneous emission, inhomogeneous broadening, and collisions in the atomic sample which may ruin the coherences in the matter waves \citep{gas}. The atoms scatter a single laser photon and recoil with a momentum of $2\hbar k$ ($k$ is the wave-vector of the incident photon) in the direction of the incident laser photon in the absence of Doppler broadening due to thermal motion of BEC atoms. The CARL instability is responsible for the exponentially enhancement in the scattered photon number and the amplitude of the density grating which arises due to the interference between the atomic wavepackets with momentum difference $2\hbar k$ \citep{moo, moo1, moo2}. The quantum mechanical illustration of the center-of-mass motion of the condensate atoms is extended theoretically from the semiclassical model of CARL \citep{pio}.  Experimentally and theoretically, the affect of motion of atoms on the superradiant scattering of light from a moving BEC was observed in \citep{bon2, fal}. The enhancement in the superradiant scattering process in the cavity of light field is possible due to the multiple reflections of the pump laser from the mirrors of the cavity which amplifies the coupling time of the atoms with the optical fields \citep{sla, mot}. The effects of a movable mirror of an optical cavity on the superradiant light scattering from a BEC in an optical lattice has been investigated \citep{bha2}.

Optomechanical cooling of a system to its quantum ground state is the main goal of the recent field of optomechanics. In the recent years, there has been an outpouring interest in the application of radiation forces to modify the centre-of-mass motion of mechanical resonators involving a vast range from macroscopic mirrors in the Laser Interferometer Gravitational Observatory (LIGO) project \citep{cor, cor1} to nano- or micromechanical cantilevers \citep{hoh, gig, arc, reg, kle, fav}, vibrating microtoroids \citep{car, sch}, membranes \citep{tho} and ultracold atoms \citep{bre, mur, bha, bha1, tre, son1, son2}. Cooling of the motion of mechanical oscillators is possible with positive radiation pressure damping whereas parametric amplification of small forces is observed with negative damping \citep{car, met}. In the fascinating work of Braginsky \citep{bra, bra1}, the mechanical damping due to radiation was first detected in the decay of an excited oscillator. Recently, the cooling of the center-of-mass motion of the mechanical oscillator i.e., both the measurement and mechanical damping of the random thermal Brownian motion, was attained by using many techniques. These involve active feedback cooling \citep{kle, coh, pog} based on position measurements, intrinsic optomechanical cooling by radiation pressure \citep{cor, gig, arc, sch, tho} or photothermal forces \citep{hoh}. A system composed of atoms in an optical cavity with moving end mirror was studied \citep{mei}. It was found that the motion of the mirror alters the dipole potential in which atoms move and the position becomes bistable.

Forty years ago, it was suggested \citep{moo3} that the relativistic motion of the mirror could create real photons from virtual photons. This phenomenon was later termed as the dynamical Casimir effect (DCE) \citep{dod}. This DCE has a common feature of creating quanta from vacuum due to the motion of macroscopic neutral boundaries or time-modulation of material properties of some macroscopic system. This fascinating phenomenon has attracted the attention of many theoreticians since decades \citep{moo3, dod, nat}. It has been observed that the DCE originates so far mostly due to the oscillation of a cavity boundary, motion of mirror in vacuum or modulation of dielectric properties of the medium inside the cavity which causes the generation of photon pairs from the electromagnetic vacuum. Recent experiments have observed the DCE which makes the problem of detecting photons generated from initial vacuum state to be realistic \citep{wil, nat, pas}. In view of recent progress, the so-called DCE has been observed with harmonic modulation of the cavity frequency in the presence of harmonic oscillators \citep{zha} and multi-level atoms \citep{dod1, dod2, dod3, dod4, dod5, dod6}. 

Given the promising developments in the field of ultracold atoms, cavity optomechanics and DCE, we propose a scheme to couple BEC to an optomechanical cavity to study the DCE. The optomechanical cavity has one cavity mirror fixed while the other mirror moving. This system is studied by considering firstly the classical motion of the movable mirror and then the quantized mirror motion. The cavity field is harmonically time-modulated in both the cases, thereby, making the system non-stationary. The main quantity of interest in the phenomenon of DCE is the number of photons produced inside the cavity which is studied here.       
   
\section{CARL-BEC Model with Classical mirror motion}

\begin{figure}[h]
\hspace{-0.0cm}
\includegraphics [scale=0.8]{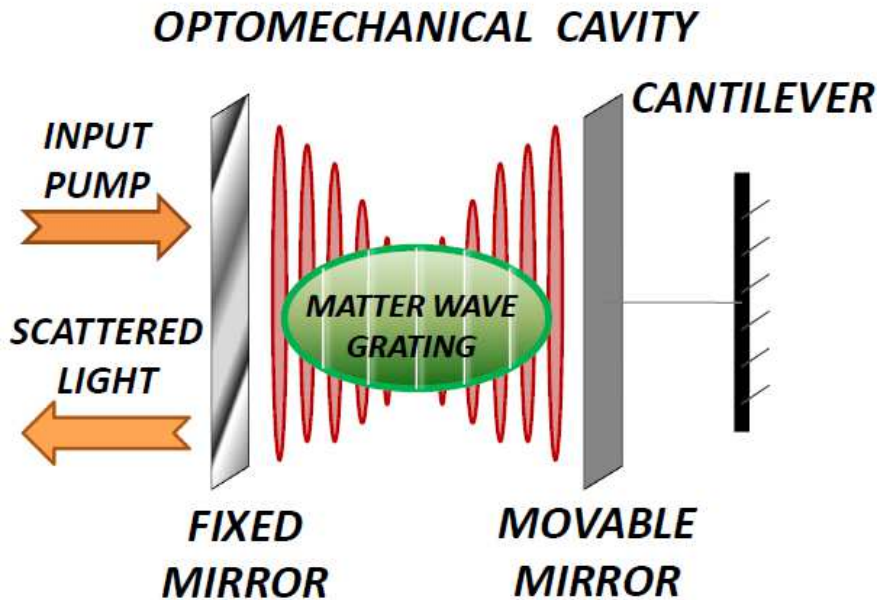}
\caption{(color online) Schematic figure of the system. The counter-propagating light fields make a periodic optical grating which further forms a periodic matter wave grating. The cavity has one mirror fixed and other movable. The fixed mirror is partially reflecting whereas the movable mirror is completely reflecting. The scattered light coming out of the fixed mirror is measured.}
\label{a}
\end{figure}

The system investigated here consists of noninteracting bosonic two-level atoms inside an optomechanical cavity coupled to two radiation fields via the electric-dipole interaction as shown in figure \ref{a}. Specifically, we take an elongated cigar shaped Bose-Einstein condensate (BEC) consisting of N two-level atoms with mass $m$ and frequency $\omega_{a}$ having transition  $|F=1>$ $\rightarrow$ $|F'=2>$. When the frequency of the harmonic trap along the axial direction is smaller than the frequency along the transverse direction then an elongated cigar shaped BEC is formed. 
The bosonic atoms are confined in optomechanical cavity whose one end mirror is fixed and other end mirror is movable. However, the mirror motion is considered to be classical. Furthermore, the cavity frequency is sinusoidally time-modulated i.e. $\omega_{c}\left(t\right)=\omega_{c}\left(1+\epsilon\sin\left(\Omega t\right)\right)$. Here, $\omega_{c}$ is the unperturbed cavity frequency, $\epsilon$ is the modulation amplitude and $\Omega$ is the modulation frequency. We consider an external pump laser with frequency $\omega_{p}$ and amplitude $\eta'$ incident on one side mirror of the optical cavity. When the longitudnal mode spacing is assumed to be larger than the induced resonance frequency shift of the cavity then one can take single longitudnal cavity mode. This approximation is considered here. The population of the atoms in the excited state is assumed to be negligible as the detuning of the pump laser and the atomic frequency is approximated to be too large as compared to the detuning of pump laser with cavity mode. Therefore, the spontaneous emission and two body dipole-dipole interactions can be safely ignored. But the two photon transitions are allowed which do not alters the atomic internal states. However, the center-of-mass motion of atom changes due to the recoil. An atom absorbs a photon from one mode and release a photon in counter-propagating mode. Therefore, it experiences a recoil kick equal to the difference between the momenta of two photons (of the order of two optical momenta). The distinct momentum states of the atomic center-of mass motion interact with two photon transition. Hence, the time-dependent Hamiltonian of the system in the dipole approximation is given as \citep{dod1}

\begin{eqnarray}\label{1}
H_{I} &=& H_{1}+H_{2}+H_{3}+H_{4}+H_{5}.
\end{eqnarray}

The free energy of photons inside the time-dependent cavity is given as

\begin{eqnarray}\label{2a}
H_{1} &=& \hbar\omega_{c}(t)a^{\dagger}a,
\end{eqnarray}

where $a (a^{\dagger})$ being the annhilation (creation) operator which obeys the commutation relation $[a,a^{\dagger}]=1$. The free energy of atoms is given by 

\begin{eqnarray}\label{2b}
H_{2} &=& \int dz\left\lbrace \psi _{g}^{\dagger}(z)\left(-\frac{\hbar ^{2}}{2m}\frac{\partial ^{2}}{\partial z^{2}}\right)\psi _{g}(z)+ \psi _{e}^{\dagger}(z)\left(-\frac{\hbar ^{2}}{2m}\frac{\partial ^{2}}{\partial z^{2}} + \hbar \omega _{a}\right)\psi _{e}(z)\right\rbrace, 
\end{eqnarray}

where $\psi_{g}(z)$, $\psi_{e}(z)$ are the ground and excited field operators of the atoms respectively. The third term in $H_{I}$ represents the atom-photon coupling which is given as

\begin{eqnarray}\label{2c}
H_{3} &=& i\hbar g_{0}a^{\dagger}\left\lbrace \int dz \psi _{g}^{\dagger}(z)\left(e^{-ikz}+e^{ikz}\right) \psi _{e}(z)\right\rbrace + H.C., 
\end{eqnarray}

where $g_{0}$ is the coupling constant. Here, the counter-propagating optical cavity field interacts with the condensate field. This counter-propagating optical cavity modes are directed towards $\pm z$ axis. Also, here $k$ represents the wavenumber of the cavity light mode. Here, the atom field operators obey the commutation relation as

\begin{eqnarray}\label{3a}
\left[\psi _{j}(z),\psi_{j'}^{\dagger}(z')\right]=\delta _{j,j'}\delta (z,z'),
\end{eqnarray}

\begin{eqnarray}\label{3b}
\left[\psi _{j}(z),\psi_{j'}(z')\right]=\left[\psi _{j}^{\dagger}(z),\psi_{j'}^{\dagger}(z')\right]=0,
\end{eqnarray}

where $j$, $j'=e,g$. Now, the energy due to external laser pump is given by the fourth term of $H_{I}$ i.e.,

\begin{eqnarray}\label{2d}
H_{4} &=& -i\hbar \eta'\left(a-a^{\dagger}\right) 
\end{eqnarray}

Here $\eta'$ is the amplitude of the external laser beam. The fifth term ($H_{5}$) in the Hamiltonian ($H_{I}$) is the squeezing term \citep{plu} given as

\begin{eqnarray}\label{2e}
H_{5}=i\hbar\chi(t)\left(a^{\dagger 2}e^{-2i\omega_{p}t}-a^{2}e^{2i\omega_{p}t}\right)
\end{eqnarray}

Due to degenerate parametric amplification of the field, this term helps in producing squeezed states of the field \citep{scu}. The photons are produced in the cavity from the vacuum state due to the fast motion of the cavity boundaries. This effect is known as dynamical Casimir effect (DCE) \citep{dod}.The effect of DCE can possibly produce squeezed states \citep{dod7, jae, sar, law, plu}. Also $\chi(t)$ is related to $\omega_{c}(t)$ as \citep{law}

\begin{equation}\label{4a}
\chi(t)=\frac{1}{4\omega_{c}(t)}\frac{d\omega_{c}(t)}{dt}.
\end{equation} 

Now we take the realistic case of small-amplitude time modulation i.e. $\vert\epsilon\vert\ll 1$. Therefore,

\begin{equation}\label{4b}
\chi(t)\approx\frac{\epsilon \Omega}{4}\cos\left(\Omega t\right)\approx2\chi_{0}\cos\left(\Omega t \right),  
\end{equation}

where, 

\begin{eqnarray}\label{4c}
\chi_{0}=\frac{\epsilon\Omega}{8}\ll 1.
\end{eqnarray}

Now, using the Heisenberg equation of motion $\left(\dot{\psi}_{e}=\left(i/\hbar\right)\left[H_{I},\psi_{e}\right]\right)$, one can adiabatically eliminate the excited state of atoms as the pump-atom detuning $\left(\Delta=\omega_{p}-\omega_{a}\right)$ is very large. This gives the Heisenberg equations of motion for the ground state field operator $\left(\psi_{g}(z)\right)$ and the cavity field annhilation operator $(a)$ by taking into account all the dissipative processes:

\begin{eqnarray}\label{5a}
\dot{\psi}_{g}=\frac{i\hbar}{2m}\frac{\partial^{2}\psi_{g}}{\partial z^{2}}-\frac{2ig_{0}^{2}a^{\dagger}a}{\Delta}\left(1+\cos(2kz)\right)\psi_{g},
\end{eqnarray}

\begin{eqnarray}\label{5b}
\dot{a}=i\Delta_{c}a-i\omega_{c}\epsilon\sin(\Omega t)a-\frac{2ig_{0}^{2}aN}{\Delta}-\frac{2ig_{0}^{2}a}{\Delta}\int dz\psi_{g}^{\dagger}\cos(2kz)\psi_{g}+2\chi(t)a^{\dagger}+\eta-\kappa a,
\end{eqnarray}

where $\Delta_{c}=\omega_{p}-\omega_{c}$, $N=\int dz\psi_{g}^{\dagger}(z)\psi_{g}(z)$ and $\kappa$ is the cavity photon decay rate. The proposed system is basically open as the photons leak from the cavity so the cavity field is damped. The second term on the right hand side of the equality in equation (\ref{5a}) represents the self consistent optical grating whose amplitude depends on time according to equation (\ref{5b}). Also, the fourth term on the right side of the equality in equation (\ref{5b}) gives the self consistent matter wave grating. 

The spatial periodic boundary conditions can be asuumed if the density of the condensate atoms is homogeneous and the optical radiation wavelength is much shorter than the length of the condensate. This leads to the ground state wavefunction of the condensate atoms as

\begin{eqnarray}\label{6}
\psi_{g}\left(z,t\right)=\sum_{n}C_{n}(t)e^{2inkz}.
\end{eqnarray}

Here, $e^{2inkz}$ represents the momentum eigenfunctions with eigenvalues $p_{z}=n(2\hbar k)$. The atomic motion considered here is based on an equivalent assumption that the atoms in a BEC are not localized inside the length of the condensate and the uncertainity in the corresponding momentum state is negligibly small. Now, the equations (\ref{5a}) and (\ref{5b}) can be reduced to the following set of ordinary differential equations

\begin{eqnarray}\label{7a}
\dot{C}_{n}=-4i\omega_{r}n^{2}C_{n}-\frac{ig_{0}^{2}NA}{\Delta}\left(2C_{n}+C_{n+1}\right),
\end{eqnarray}

\begin{eqnarray}\label{7b}
\dot{C}_{n+1}=-4i\omega_{r}(n+1)^{2}C_{n+1}-\frac{ig_{0}^{2}NA}{\Delta}\left(2C_{n+1}+C_{n}\right),
\end{eqnarray}

\begin{eqnarray}\label{7c}
\dot{a}'=i\Delta_{c}a'-i\omega_{c}\epsilon\sin(\Omega t)a'-\frac{2ig_{0}^{2}a'N}{\Delta}-\frac{2ig_{0}^{2}a'N}{\Delta}C_{n+1}^{*}C_{n}+2\chi(t){a'}^{\dagger}+\eta-\kappa a'.
\end{eqnarray}

Here, $A={a'}^{\dagger} a'$, $n$ is the initial momentum level, $n+1$ is the final momentum level and $\omega_{r}=\hbar k^{2}/2m$ is the recoil frequency. Also, here $a'$ is the rescaled photon operator with $a\rightarrow a'\sqrt{N}$ and $\eta$ is the rescaled external pump amplitude with $\eta\rightarrow\eta\sqrt{N}$. 

Defining coherence  $S=C_{n}C_{n+1}^{*}$ and the population difference between the two states as $W=\vert C_{n}\vert ^{2}-\vert C_{n+1}\vert ^{2}$, we obtain the following equations of motion 

\begin{eqnarray}\label{8a}
\dot{S}=4i\omega_{r}\left(1+2n\right)S+\frac{ig_{0}^{2}NA}{\Delta}W-\gamma S,
\end{eqnarray}

\begin{eqnarray}\label{8b}
\dot{W}=\frac{2ig_{0}^{2}NA}{\Delta}\left(S-S^{*}\right),
\end{eqnarray}

\begin{eqnarray}\label{8c}
\dot{a}'=i\Delta_{c}a'-i\omega_{c}\epsilon\sin(\Omega t)a'-\frac{2ig_{0}^{2}a'N}{\Delta}-\frac{2ig_{0}^{2}a'N}{\Delta}S+2\chi(t){a'}^{\dagger}+\eta-\kappa a'.
\end{eqnarray}

Here, we have introduced a damping term in equation (\ref{8a}) to account for the decay of the coherence between the two motional states $n$ and $n+1$. Now separating the above equations of motion into real and imaginary parts, we get

\begin{eqnarray}\label{9a}
\dot{S}_{r}=-4\omega_{r}(1+2n)S_{i}-\gamma S_{r},
\end{eqnarray}

\begin{eqnarray}\label{9b}
\dot{S}_{i}=4\omega_{r}(1+2n)S_{r}+g\left(a_{r}^{2}+a_{i}^{2}\right)W-\gamma S_{i},
\end{eqnarray}

\begin{eqnarray}\label{9c}
\dot{W}=-4g\left(a_{r}^{2}+a_{i}^{2}\right)S_{i},
\end{eqnarray}

\begin{eqnarray}\label{9d}
\dot{a}_{r}=-\Delta_{c}a_{i}+\omega_{c}\epsilon\sin(\Omega t)a_{i}+2ga_{i}+2g\left(a_{i}S_{r}+a_{r}S_{i}\right)+2\chi(t)a_{r}+\eta-\kappa a_{r},
\end{eqnarray}

\begin{eqnarray}\label{9e}
\dot{a}_{i}=\Delta_{c}a_{r}-\omega_{c}\epsilon\sin(\Omega t)a_{r}-2ga_{r}-2g\left(a_{r}S_{r}-a_{i}S_{i}\right)-2\chi(t)a_{i}-\kappa a_{i},
\end{eqnarray}

where, the renormalized coupling constant is given as $g=g_{0}^{2}N/\Delta$. We investigate the system in the quantum superradiant regime $g/\omega_{r}\sqrt{N}<2\sqrt{\kappa/\omega_{r}}$. In this regime, the condensate momentum alters by $2\hbar k$ as each atom scatters a single photon only. As soon as the pump power builds up in the ring cavity, the collective dynamics start. We solve the above coupled differential equations of motion (\ref{9a}-\ref{9e}) using Mathematica 9.0 to study this dynamics via the evolution of the power $A(t)={a'}^{\dagger}a'$. We further identify two more interesting regimes under which the system is investigated. First is strong modulation $\left(\chi_{0}\gtrsim g\right)$ and other is weak modulation $\left(\chi_{0}\ll g\right)$. 

\begin{figure}[h]
\hspace{-0.0cm}
\begin{tabular}{cc}
\includegraphics [scale=0.75]{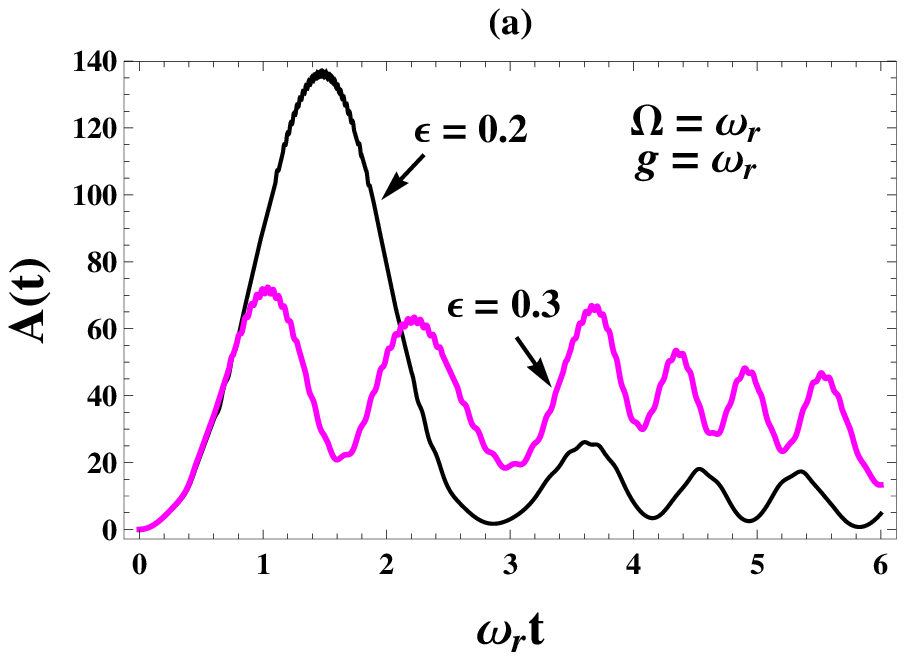}& \includegraphics [scale=0.75] {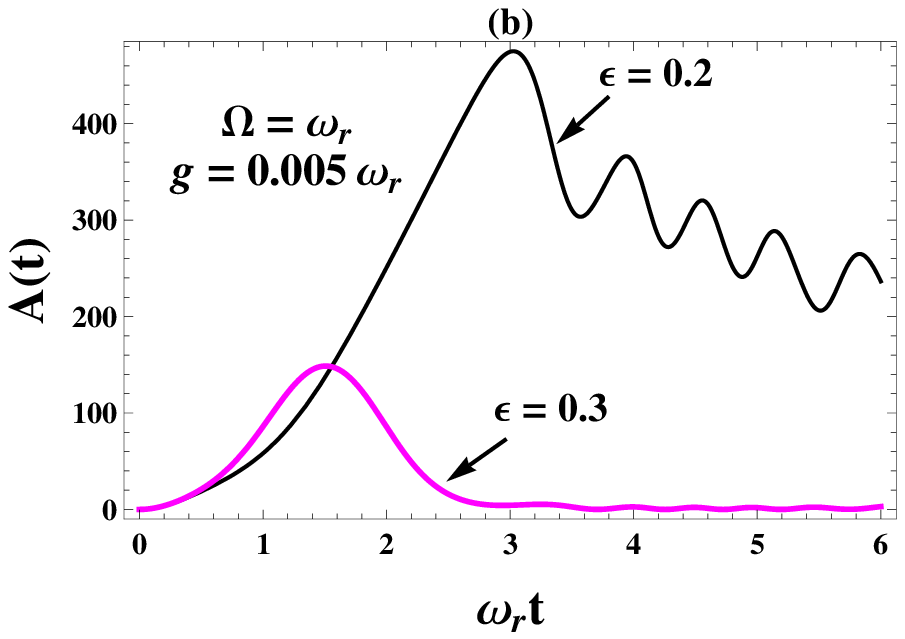}\\
 \end{tabular}
\caption{(Color online) Time signal of the scattered light power $\left(A(t)=\langle a^{\dagger}a\rangle\right)$ with classical motion of the mirror using time modulated cavity frequency at resonant frequency $\left(\Omega = \omega_{r}\right)$ for two modulation amplitudes $\epsilon = 0.2$ (thin line) and $\epsilon = 0.3$ (thick line). Parameters used are $n=0$, $\gamma=0.1\omega_{r}$, $\Delta_{c}=5\omega_{r}$, $\omega_{c}=28.8\omega_{r}$, $\eta=10\omega_{r}$ and $\kappa=0.1\omega_{r}$. For plot (a) Weak modulation $\left(g=\omega_{r}\right)$ and plot (b) Strong modulation $\left(g=0.005\omega_{r}\right)$.} 
\label{b}
\end{figure}

Now, the transition from the Bragg scattering regime to the quantum superradiant regime has been investigated. We have considered the BEC to be at rest initially $\left(n=0\right)$. Figure (\ref{b}) shows the evolution of power $A(t)$ versus scaled time $(\omega_{r} t)$ by considering the classical motion of mirror. The plot in Fig. (\ref{b}(a)) corresponds to the weak modulation $\left(g=\omega_{r}\right)$  with two modulation amplitudes $\epsilon=0.2$ (thin line) and $\epsilon=0.3$ (thick line). The plot shows damped periodic oscillations with time. These oscillations depict the typical Bragg resonances due to two-photon transitions between the momentum states. Under the strong modulation $\left(g=0.005\omega_{r}\right)$ as shown in Fig. (\ref{b}(b)), the Bragg oscillations vanish with time for higher modulation amplitude $(\epsilon=0.3)$. Therefore, one can switch between the pure superradiant regime and Bragg scattering regime by varying the modulation amplitude. Reducing the renormailzed coupling constant $(g)$, increases the effect of DCE and decreases the Bragg scattering superradiant process. One more thing noticeable here is that the amplitude of oscillations is very high under the strong modulation as compared to weak modulation. This is because under the strong modulation the effect of DCE is dominant which produces extra photons in the cavity from the vacuum state. It can be further observed that, initially, the amplitude of number of scattered photon is high for lower modulation amplitude as compared to higher modulation amplitude. This is due to the fact that as we are increasing the modulation amplitude, the dispersive effect increases. This reduces the matter wave grating leading to less scattering of photons in the cavity. In the next section, we consider the same system with quantized motion of the movable mirror.  

\section{CARL-BEC Model with Quantized mirror motion}

In this section, we consider the similar system with BEC confined in an optomechanical cavity. In addition to this, here we consider the quantized motion of the movable mirror of the cavity. The movable mirror is treated as a harmonic oscillator which oscillates with frequency $\omega_{m}$ and has mass $m$. Hence again, the cavity frequency is harmonically modulated due to the rapid motion of the mirror. A force proportional to the photon number acts on the movable mirror. As a result of this, various couplings of the system also modulate sinusoidally with time. The Hamiltonian with quantized motion of the mirror under rotating-wave and dipole approximation is given as

\begin{eqnarray}\label{10}\nonumber
H_{II} &=& \hbar\omega_{c}a^{\dagger}a+\int dz\left\lbrace \psi _{g}^{\dagger}(z)\left(-\frac{\hbar ^{2}}{2m}\frac{\partial ^{2}}{\partial z^{2}}\right)\psi _{g}(z)+ \psi _{e}^{\dagger}(z)\left(-\frac{\hbar ^{2}}{2m}\frac{\partial ^{2}}{\partial z^{2}} + \hbar \omega _{a}\right)\psi _{e}(z)\right\rbrace\nonumber \\ &+& i\hbar g_{0}a^{\dagger}\left\lbrace \int dz \psi _{g}^{\dagger}(z)\left(e^{-ikz}+e^{ikz}\right) \psi _{e}(z)\right\rbrace -i\hbar g_{0}a\left\lbrace \int dz \psi _{e}^{\dagger}(z)\left(e^{ikz}+e^{-ikz}\right) \psi _{g}(z)\right\rbrace\nonumber \\ &-& i\hbar \eta'\left(a-a^{\dagger}\right)+i\hbar\chi(t)\left(a^{\dagger 2}e^{-2i\omega_{p}t}-a^{2}e^{2i\omega_{p}t}\right)\left(b+b^{\dagger}\right)+\hbar\omega_{m}b^{\dagger}b\nonumber \\ &+& \hbar k_{0}(t)a^{\dagger}a\left(b+b^{\dagger}\right)+i\hbar G(t)a^{\dagger}\left(b+b^{\dagger}\right)\left\lbrace \int dz \psi _{g}^{\dagger}(z)\left(e^{-ikz}+e^{ikz}\right) \psi _{e}(z)\right\rbrace\nonumber \\ &-& i\hbar G(t)a\left(b+b^{\dagger}\right)\left\lbrace \int dz \psi _{e}^{\dagger}(z)\left(e^{ikz}+e^{-ikz}\right) \psi _{g}(z)\right\rbrace .
\end{eqnarray} 

The above Hamiltonian is derived in Appendix A. In addition to the terms in in Hamiltonian $(H_{I})$, there are some extra terms in the Hamiltonian $(H_{II})$ as the quantized motion of the mirror is considered here. Here the squeezing term (sixth term in eqn. (\ref{10})) is modified and is dependent on the coupling between the intracavity photons and position of the movable mirror. The annhilation (creation) operator of the mechanical mode is given by $b (b^{\dagger})$. The seventh term gives the free energy of the mechanical oscillator. The time dependent coupling between the intensity of cavity light field and mechanical mode of the mirror is represented by the eighth term in eqn. (\ref{10}) where $k_{0}(t)$ is the corresponding time dependent coupling parameter which is defined as $k_{0}(t)=\omega_{c}\epsilon\sin (\Omega t)$. The ninth term describes the three body time dependent interaction between the condensate mode, counter propagating cavity optical mode and mechanical mode of the movable mirror. The last term in $H_{II}$ is the hyper conjugate of the ninth term. Here $G(t)$ is the time dependent three body coupling parameter which is given as $G(t)=g_{0}\epsilon\sin (\Omega t)/2$. Again we eliminate the excited state of condensate atoms. Therefore, the Heisenberg equations of motion of the system for all the degrees of system becomes 

\begin{eqnarray}\label{11a}\nonumber
\dot{\psi}_{g} &=& \frac{i\hbar}{2m}\frac{\partial^{2}\psi_{g}}{\partial z^{2}}-\frac{2ig_{0}a^{\dagger}a}{\Delta}\left(g_{0}+G(t)\left(b+b^{\dagger}\right)\right)\left(1+\cos(2kz)\right)\psi_{g}\nonumber \\ &-& \frac{2iG(t)a^{\dagger}a}{\Delta}\left(b+b^{\dagger}\right)\left(g_{0}+G(t)\left(b+b^{\dagger}\right)\right)\left(1+\cos(2kz)\right)\psi_{g},
\end{eqnarray}

\begin{eqnarray}\label{11b}\nonumber
\dot{a} &=& i\Delta_{c}a-ik_{0}(t)a \left(b+b^{\dagger}\right)-\frac{2ig_{0}aN}{\Delta}\left(g_{0}+G(t)\left(b+b^{\dagger}\right)\right)\nonumber \\ &-& \frac{2ig_{0}a}{\Delta}\left(g_{0}+G(t)\left(b+b^{\dagger}\right)\right)\int dz\psi_{g}^{\dagger}\cos(2kz)\psi_{g}+2\chi(t)a^{\dagger}\left(b+b^{\dagger}\right)\nonumber \\ &+& \eta-\frac{2iG(t)aN}{\Delta}\left(b+b^{\dagger}\right)\left(g_{0}+G(t)\left(b+b^{\dagger}\right)\right)\nonumber \\ &-& \frac{2iG(t)a}{\Delta}\left(b+b^{\dagger}\right)\left(g_{0}+G(t)\left(b+b^{\dagger}\right)\right)\int dz\psi_{g}^{\dagger}\cos(2kz)\psi_{g},
\end{eqnarray}

\begin{eqnarray}\label{11c}\nonumber
\dot{b} &=& -i\omega_{m}b-ik_{0}(t)a^{\dagger}a-\frac{2iG(t)\left(1+2a^{\dagger}a\right)N}{\Delta}\left(g_{0}+G(t)\left(b+b^{\dagger}\right)\right)\nonumber \\ &-& \frac{2iG(t)\left(1+2a^{\dagger}a\right)}{\Delta}\left(g_{0}+G(t)\left(b+b^{\dagger}\right)\right)\int dz\psi_{g}^{\dagger}\cos(2kz)\psi_{g}+\chi(t)\left({a^{\dagger}}^{2}-a^{2}\right).
\end{eqnarray}

In eqn. (\ref{11b}), the second term represents the radiation pressure coupling of the cavity optical mode with the mirror displacement. As done in previous section, here also we use the ground state wavefunction of the condensate atom given by eqn. (\ref{6}). Therefore, by considering all the dissipation processes of the system we get the following set of reduced differential equations of motion

\begin{eqnarray}\label{12a}
\dot{C}_{n}=-4i\omega_{r}n^{2}C_{n}-\frac{iNA}{\Delta}\left(g_{0}+G(t)x_{m}\right)^{2}\left(2C_{n}+C_{n+1}\right),
\end{eqnarray}

\begin{eqnarray}\label{12b}
\dot{C}_{n+1}=-4i\omega_{r}(n+1)^{2}C_{n+1}-\frac{iNA}{\Delta}\left(g_{0}+G(t)x_{m}\right)^{2}\left(2C_{n+1}+C_{n}\right),
\end{eqnarray}

\begin{eqnarray}\label{12c}\nonumber
\dot{a'} &=& \left(i\Delta_{c}-ik_{0}(t)x_{m}\right)a'-\frac{2ia'N}{\Delta}\left(g_{0}+G(t)x_{m}\right)^{2}\nonumber \\ &-& \frac{2ia'N}{\Delta}\left(g_{0}+G(t)x_{m}\right)^{2}C_{n+1}^{*}C_{n}+\eta +2\chi(t){a'}^{\dagger}x_{m}-\kappa a',
\end{eqnarray}

\begin{eqnarray}\label{12d}
\dot{x}_{m}=\omega_{m}p_{m},
\end{eqnarray}

\begin{eqnarray}\label{12e}\nonumber
\dot{p}_{m} &=& -\omega_{m}x_{m}-2k_{0}(t)NA-2i\chi(t)N\left({a'}^{\dagger 2}-{a'}^{2}\right)-\gamma_{m}p_{m}\nonumber \\ &-& \frac{4G(t)N^{2}}{\Delta}\left(1+2A\right)\left(g_{0}+G(t)x_{m}\right)\left(1+C_{n}^{*}C_{n+1}+C_{n+1}^{*}C_{n}\right),
\end{eqnarray}

where $\gamma_{m}$ is the dissipation rate of the mechanical mode, $x_{m}=\left(b+b^{\dagger}\right)$ is the position of the movable mirror and $p_{m}=i\left(b^{\dagger}-b\right)$ is the momentum of the movable mirror. All the other assumptions are same as in the previous case.

In equation (\ref{12c}), the first term on the right hand side of the equality gives the renormalized cavity detuning. So the cavity detuning is altering with time due to $k_{0}(t)$. One should note that when $k_{0}(t)=0$ (stationary mirror) and $\Delta_{c}=0$ gives the Bragg condition of the scattering. This condition arises due to the energy and momentum conservation. We also noticed from the equation (\ref{12c}) that the Bragg resonances are effcted by dynamical dispersive effect of the mirror motion which is proportional on the mirror displacement $x_{m}$. In the linear regime, when $x_{m}=x_{0}$ (some initial value) and $a$ is still small, the Bragg condition becomes $\Delta_{c}=k_{0}(t)$. As defined earlier, the equations of motion of the coherences and other equations in terms of coherences are given as

\begin{eqnarray}\label{13a}
\dot{S}=4i\omega_{r}\left(1+2n\right)S+\frac{iNA}{\Delta}\left(g_{0}+G(t)x_{m}\right)^{2}W-\gamma S,
\end{eqnarray}

\begin{eqnarray}\label{13b}
\dot{W}=\frac{2iNA}{\Delta}\left(g_{0}+G(t)x_{m}\right)^{2}\left(S-S^{*}\right),
\end{eqnarray}

\begin{eqnarray}\label{13c}\nonumber
\dot{a'} &=& i\Delta_{c}a'-ik_{0}(t)a'x_{m}-\frac{2ia'N}{\Delta}\left(g_{0}+G(t)x_{m}\right)^{2}\nonumber \\ &-& \frac{2ia'N}{\Delta}\left(g_{0}+G(t)x_{m}\right)^{2}S+\eta +2\chi(t){a'}^{\dagger}x_{m}-\kappa a',
\end{eqnarray}

\begin{eqnarray}\label{13d}
\dot{x}_{m}=\omega_{m}p_{m},
\end{eqnarray}

\begin{eqnarray}\label{13e}\nonumber
\dot{p}_{m} &=& -\omega_{m}x_{m}-2k_{0}(t)NA-2i\chi(t)N\left({a'}^{\dagger 2}-{a'}^{2}\right)-\gamma_{m}p_{m}\nonumber \\ &-& \frac{4G(t)N^{2}}{\Delta}\left(1+2A\right)\left(g_{0}+G(t)x_{m}\right)\left(1+S+S^{*}\right).
\end{eqnarray}

\begin{figure}[h]
\hspace{-0.0cm}
\begin{tabular}{cc}
\includegraphics [scale=0.75]{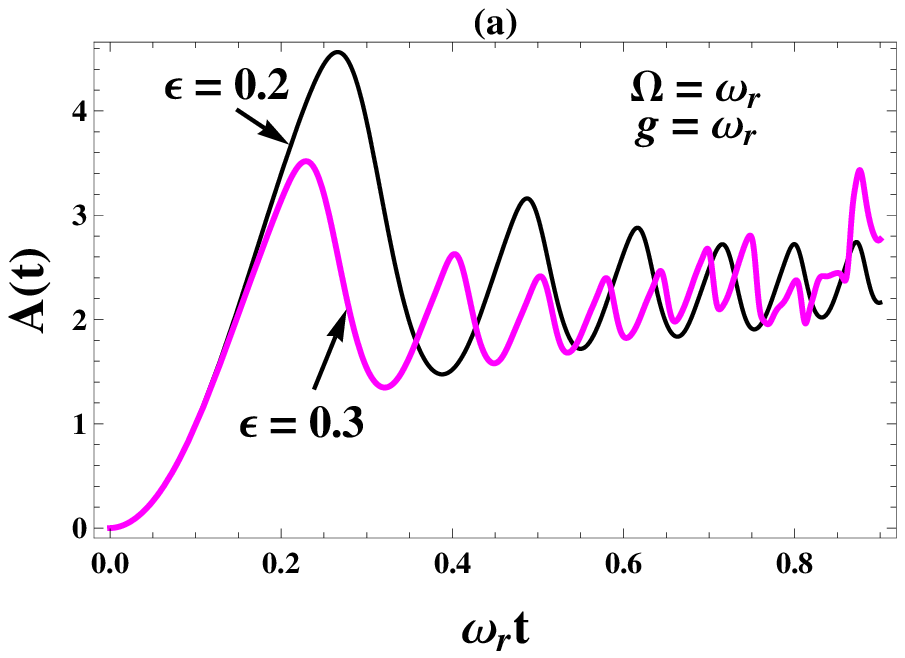}& \includegraphics [scale=0.75] {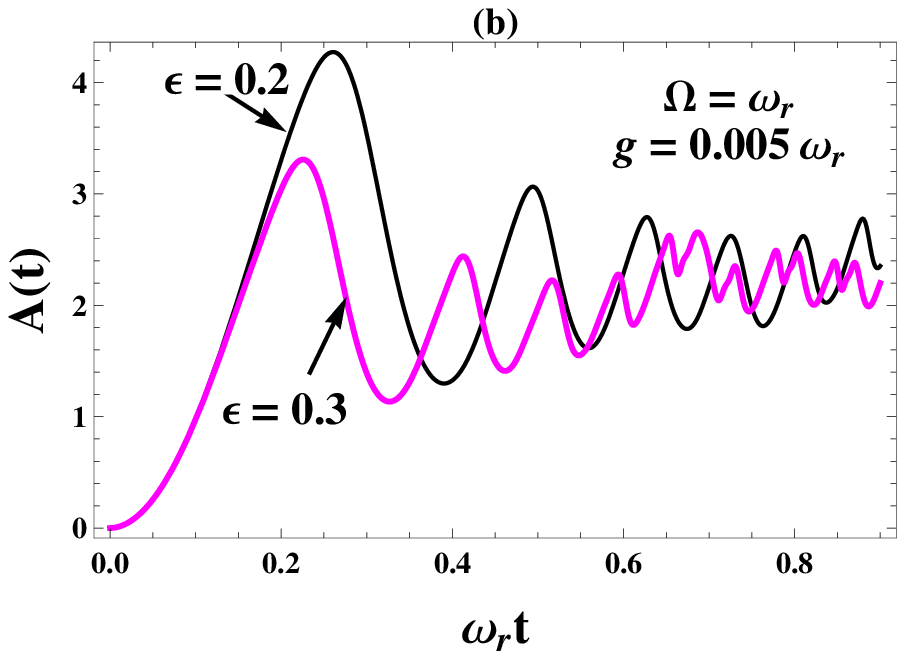}\\
\includegraphics [scale=0.75]{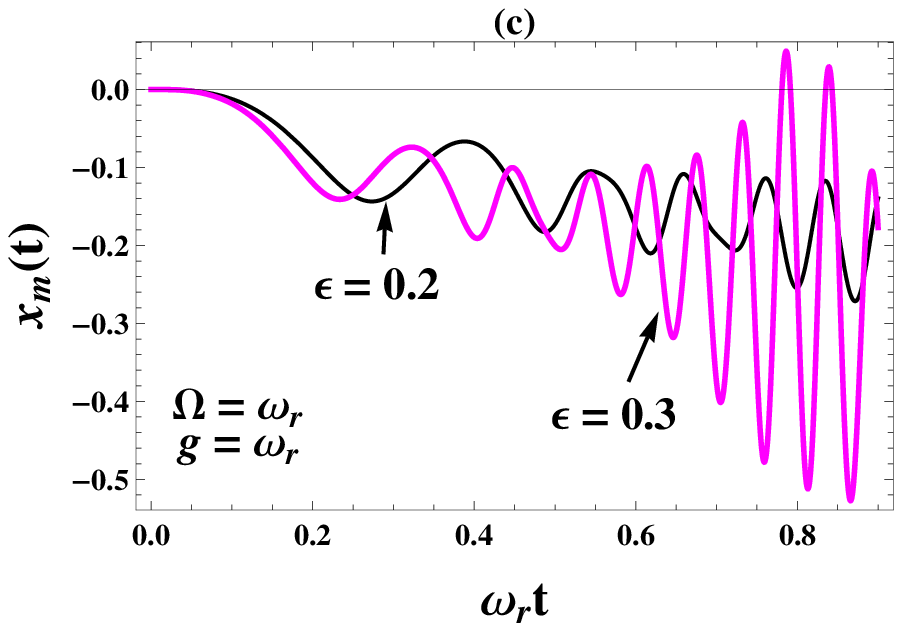}& \includegraphics [scale=0.75] {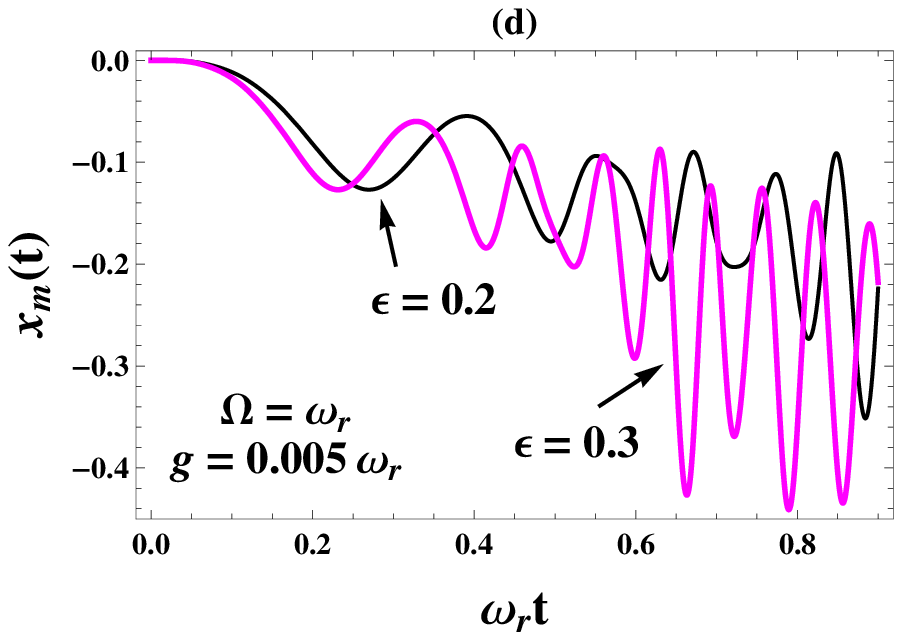}\\
 \end{tabular}
\caption{(Color online) Plots (a) and (b) show time signal of the scattered light power $\left(A(t)=\langle a^{\dagger}a\rangle\right)$ and plots (c) and (d) show the corresponding position dynamics of the movable mirror $\left(x_{m}(t)\right)$ versus scaled time $\left(\omega_{r}t\right)$ with quantized motion of the mirror using time modulated cavity frequency at resonant frequency $\left(\Omega = \omega_{r}\right)$ for two modulation amplitudes $\epsilon = 0.2$ (thin line) and $\epsilon = 0.3$ (thick line). Parameters used are $n=0$, $\gamma=0.1\omega_{r}$, $\Delta_{c}=5\omega_{r}$, $\omega_{c}=28.8\omega_{r}$, $\eta=10\omega_{r}$, $\kappa=0.1\omega_{r}$, $\omega_{m}=100\omega_{r}$, $N=100$ and $\gamma_{m}=\omega_{r}$. For plots (a) and (c) Weak modulation $\left(g=\omega_{r}\right)$ and plots (b) and (d) Strong modulation $\left(g=0.005\omega_{r}\right)$.} 
\label{c}
\end{figure}

\begin{figure}[h]
\hspace{-0.0cm}
\begin{tabular}{cc}
\includegraphics [scale=0.75]{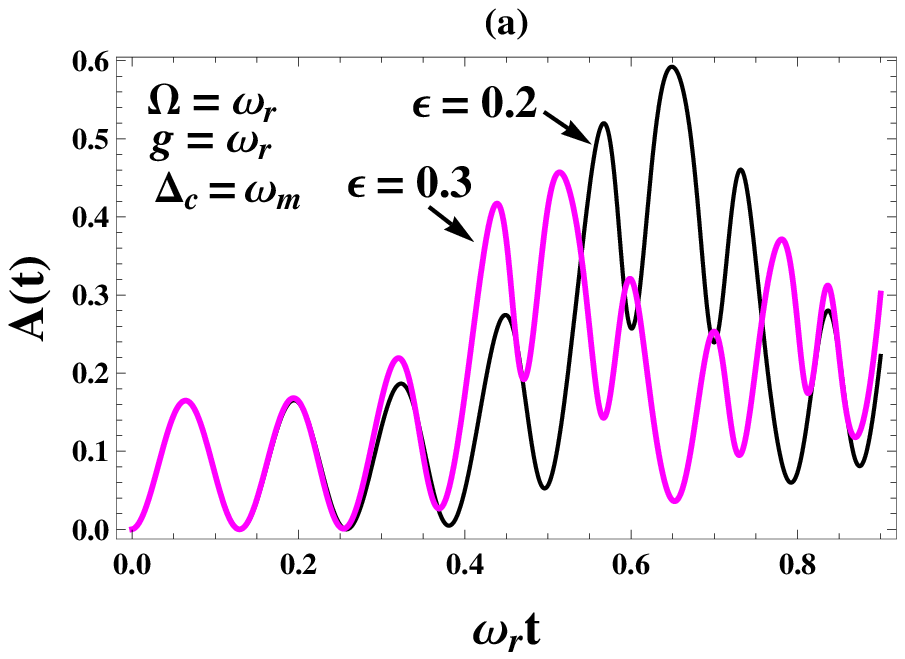}& \includegraphics [scale=0.75] {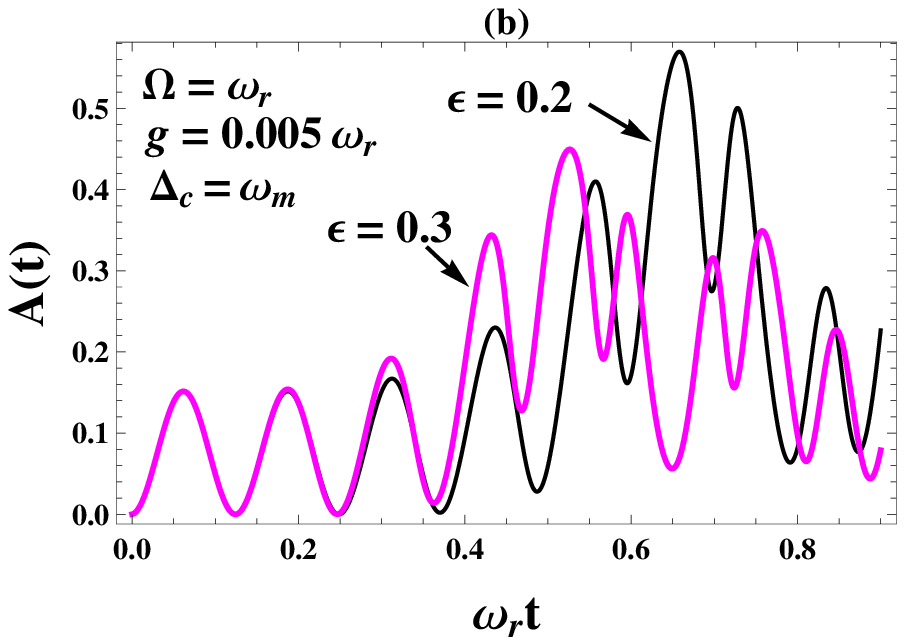}\\
\includegraphics [scale=0.75]{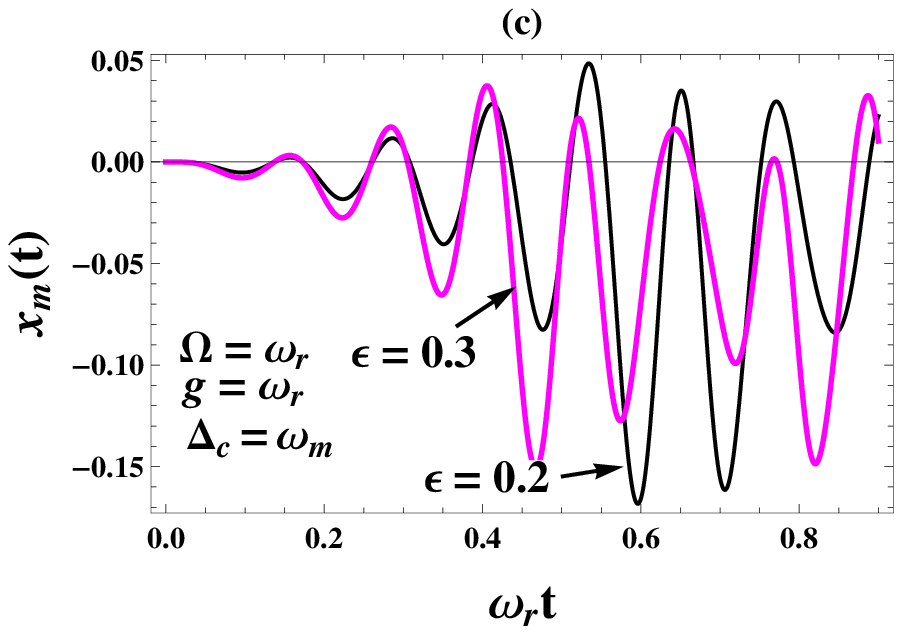}& \includegraphics [scale=0.75] {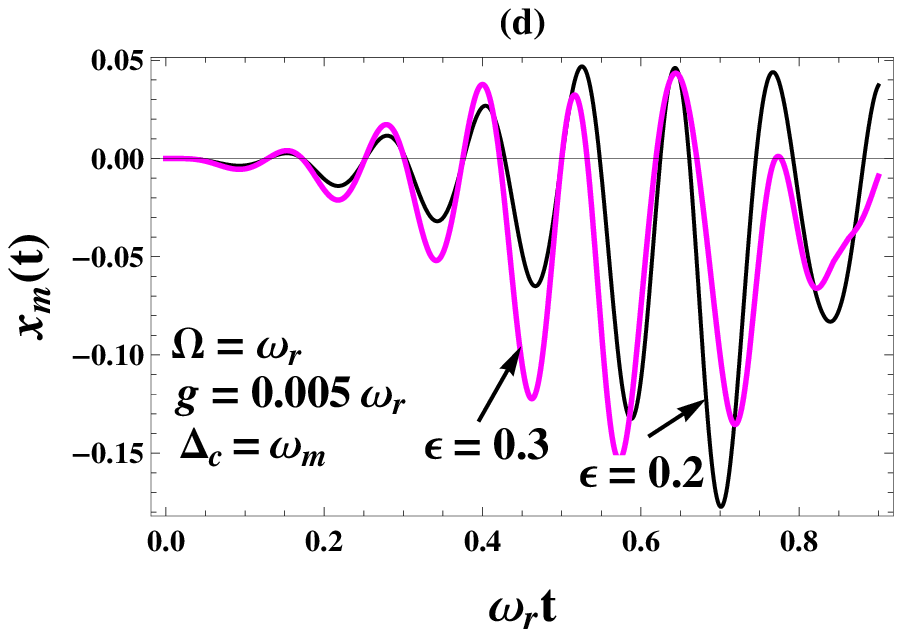}\\
 \end{tabular}
\caption{(Color online) Plots (a) and (b) show the influence of cavity detuning on the time signal of the scattered light power $\left(A(t)=\langle a^{\dagger}a\rangle\right)$ and plots (c) and (d) show the corresponding position dynamics of the movable mirror $\left(x_{m}(t)\right)$ using the quantized motion of the movable mirror with time modulated cavity frequency at resonant frequency $\left(\Omega = \omega_{r}\right)$ for two modulation amplitudes $\epsilon = 0.2$ (thin line) and $\epsilon = 0.3$ (thick line) with $\Delta_{c}=\omega_{m}$ where $\omega_{m}=50\omega_{r}$. Plots (a) and (c) show the variation of $A(t)$ and $x_{m}(t)$ respectively under weak modulation $\left(g=\omega_{r}\right)$. Plots (b) and (d) show the same variation under strong modulation $\left(g=0.005\omega_{r}\right)$. Other parameters used are same as figure (\ref{c}).} 
\label{d}
\end{figure}

\begin{figure}[h]
\hspace{-0.0cm}
\begin{tabular}{cc}
\includegraphics [scale=0.75]{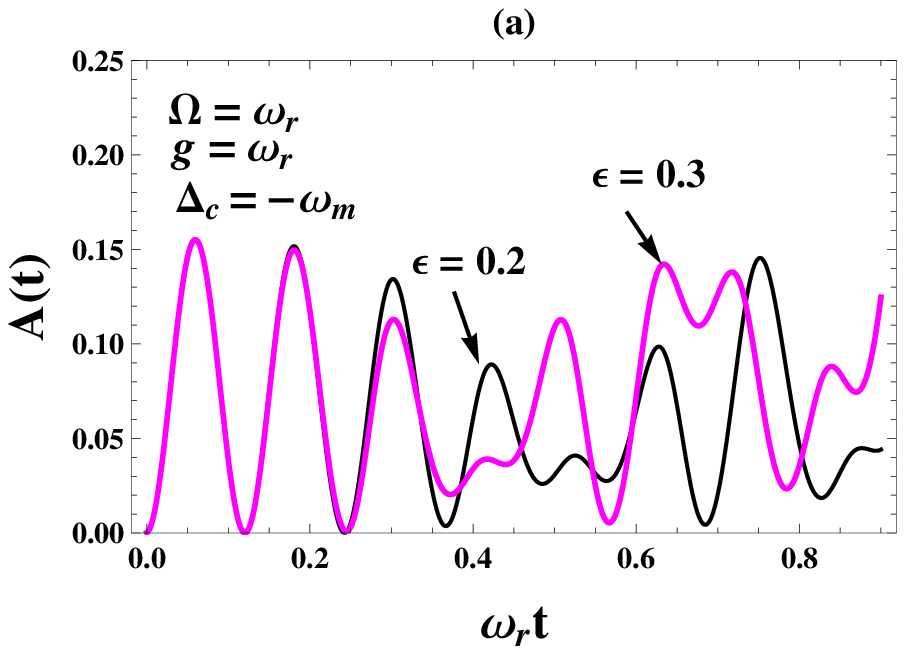}& \includegraphics [scale=0.75] {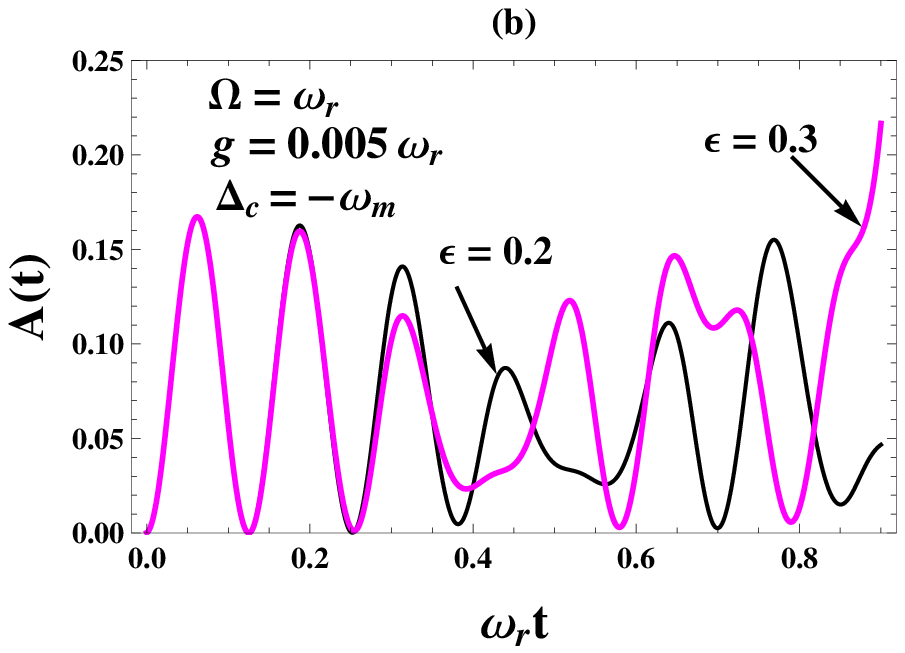}\\
\includegraphics [scale=0.75]{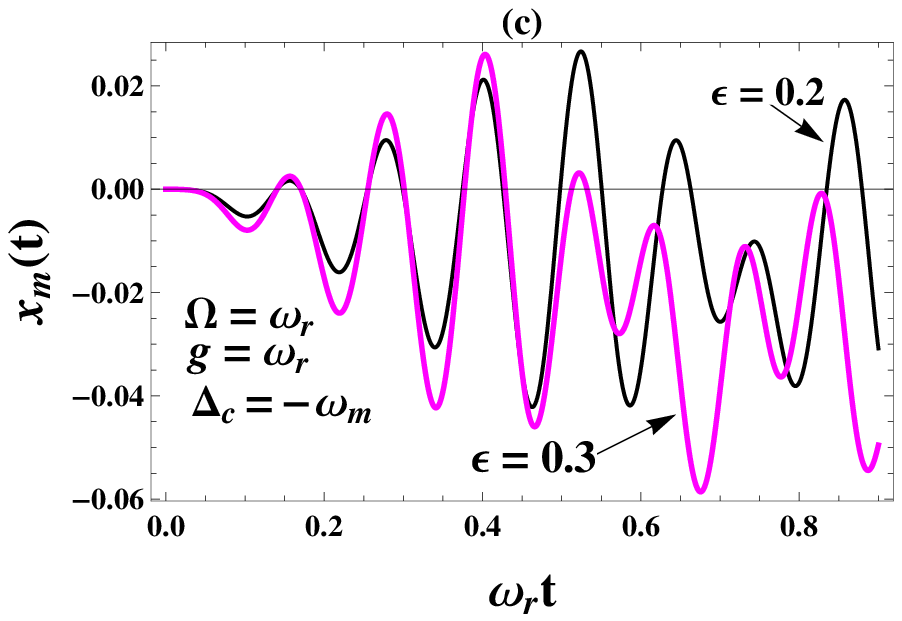}& \includegraphics [scale=0.75] {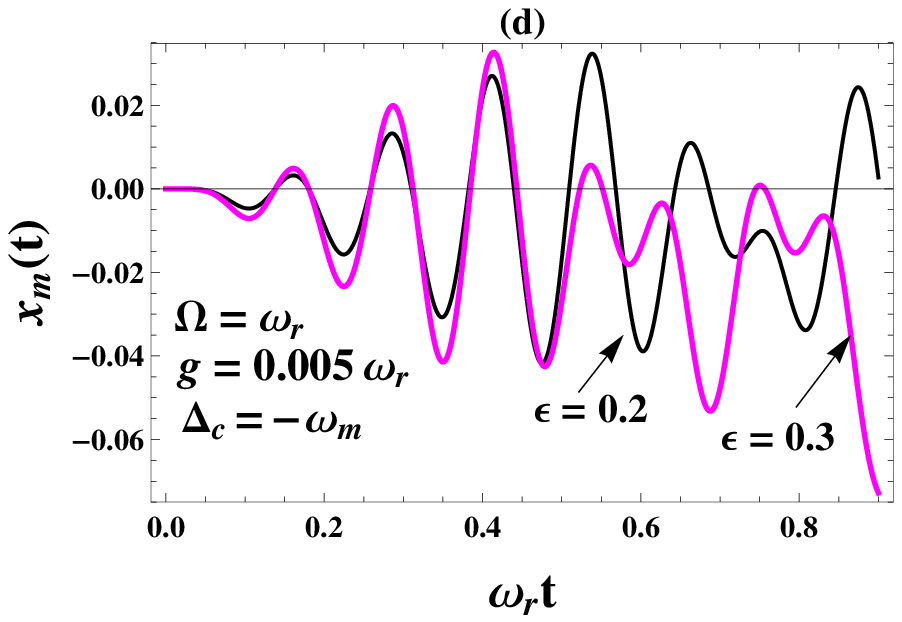}\\
 \end{tabular}
\caption{(Color online) Plots (a) and (b) show the influence of cavity detuning on the time signal of the scattered light power $\left(A(t)=\langle a^{\dagger}a\rangle\right)$ and plots (c) and (d) show the corresponding position dynamics of the movable mirror $\left(x_{m}(t)\right)$ using the quantized motion of the movable mirror with time modulated cavity frequency at resonant frequency $\left(\Omega = \omega_{r}\right)$ for two modulation amplitudes $\epsilon = 0.2$ (thin line) and $\epsilon = 0.3$ (thick line) with $\Delta_{c}=-\omega_{m}$. Plots (a) and (c) show the variation of $A(t)$ and $x_{m}(t)$ respectively under weak modulation $\left(g=\omega_{r}\right)$. Plots (b) and (d) show the same variation under strong modulation $\left(g=0.005\omega_{r}\right)$. Other parameters used are same as figure (\ref{c}).} 
\label{e}
\end{figure}

Now again separating the above set of differential equations into real and imaginary parts, we obtain the following equations

\begin{eqnarray}\label{14a}
\dot{S}_{r}=-4\omega_{r}\left(1+2n\right)S_{i}-\gamma S_{r},
\end{eqnarray}

\begin{eqnarray}\label{14b}
\dot{S}_{i}=4\omega_{r}\left(1+2n\right)S_{r}+\frac{NA}{\Delta}\left(g_{0}+G(t)x_{m}\right)^{2}W-\gamma S_{i},
\end{eqnarray}

\begin{eqnarray}\label{14c}
\dot{W}=-\frac{4NA}{\Delta}\left(g_{0}+G(t)x_{m}\right)^{2}S_{i},
\end{eqnarray}

\begin{eqnarray}\label{14d}\nonumber
\dot{a}_{r} &=& -\Delta_{c}a_{i}+k_{0}(t)a_{i}x_{m}+\frac{2a_{i}N}{\Delta}\left(g_{0}+G(t)x_{m}\right)^{2}\nonumber \\ &+& \frac{2N}{\Delta}\left(g_{0}+G(t)x_{m}\right)^{2}\left(a_{r}S_{i}+a_{i}S_{r}\right)+\eta +2\chi(t)a_{r}x_{m}-\kappa a_{r},
\end{eqnarray}

\begin{eqnarray}\label{14e}\nonumber
\dot{a}_{i} &=& \Delta_{c}a_{r}-k_{0}(t)a_{r}x_{m}-\frac{2a_{r}N}{\Delta}\left(g_{0}+G(t)x_{m}\right)^{2}\nonumber \\ &-& \frac{2N}{\Delta}\left(g_{0}+G(t)x_{m}\right)^{2}\left(a_{r}S_{r}-a_{i}S_{i}\right)-2\chi(t)a_{i}x_{m}-\kappa a_{i},
\end{eqnarray}

\begin{eqnarray}\label{14f}
\dot{x}_{m}=\omega_{m}p_{m},
\end{eqnarray}

\begin{eqnarray}\label{14g}\nonumber
\dot{p}_{m} &=& -\omega_{m}x_{m}-2k_{0}(t)NA-2i\chi(t)N\left(a_{r}a_{i}+a_{i}a_{r}\right)-\gamma_{m}p_{m}\nonumber \\ &-& \frac{4G(t)N^{2}}{\Delta}\left(1+2A\right)\left(g_{0}+G(t)x_{m}\right)\left(1+2S_{r}\right).
\end{eqnarray}

Now we renormalize $x_{m}\rightarrow Nx_{m}$ and $p_{m}\rightarrow Np_{m}$ and solve the above coupled differential equations of motion using Mathematica 9.0. Again we investigate the system under the same quantum superradiant regime. Also, the system is studied under the same limits as in previous section. Apart from the observations noticed in previous section, here we note some additional features as described below.   

Figures (\ref{c} (a)) and (\ref{c}(b)) show the time signal of the scattered light power $(A(t))$ and figures (\ref{c}(c)) and (\ref{c}(d)) show the corresponding position dynamics of the movable mirror $\left(x_{m}(t)\right)$ versus scaled time $\left(\omega_{r}t\right)$ using time modulated cavity frequency at resonant modulating frequency $\left(\Omega = \omega_{r}\right)$ for two modulation amplitudes $\epsilon = 0.2$ (thin line) and $\epsilon = 0.3$ (thick line). The plots demonstrate an initial rise in the amplitude of scattered light (see figs. (\ref{c}(a)) and (\ref{c}(b))) while an initial dip in the position dynamics of the mirror (see figs. (\ref{c}(c)) and (\ref{c}(d))). As time passes, the oscillations of the scattered light is damped whereas the corresponding oscillations of the position dynamics of the movable mirror is enhanced. This clearly shows the exchange of energy between different degrees of freedom of the system. Also, as observed with the classical motion of the mirror, here initially the amplitude of the number of scattered photons is more for lower modulation amplitude as compared to higher modulation amplitude and the amplitude of the corresponding mirror displacement is less for lower modulation amplitude as compared to higher modulation amplitude. As compared to the case of classical mirror motion (see fig. \ref{b}), here the amplitude of the number of scattered photons is less. This is because here there is an additional degree of freedom (movable mirror). So here the exchange of energy takes place with an extra degree of freedom which decreases the scattered photons in the cavity. Also, in the present case, there is not much difference in the plots of number of scattered photons under the strong and weak modulation in contrast to the case of classical motion of the mirror. This means that the effect of modulating the cavity frequency doesnot alter much the number of scattered photons in the cavity by considering the quantized mirror motion.  

Plot (\ref{d}) shows the influence of cavity detuning on the time signal of the scattered light power $\left(A(t)\right)$ and the corresponding position dynamics of the movable mirror at resonant modulating frequency $\left(\Omega = \omega_{r}\right)$ for two modulation amplitudes $\epsilon = 0.2$ (thin line) and $\epsilon = 0.3$ (thick line) with $\Delta_{c}=\omega_{m}$. Plots (\ref{d}(a)) and (\ref{d}(b)) illustrate amplification in the scattered light. This is due to the DCE which produces many photons inside the cavity from the vacuum state, leading to enhancement in the number of scattered photons in the cavity. Plots (\ref{d}(c)) and (\ref{d}(d)) show the corresponding suppression in the position dynamics of the movable mirror. This variation is again due to the balance of energy between different degrees of freedom. Plot (\ref{e}) illustates the same influence of cavity detuning as in figure (\ref{d}) with $\Delta_{c}=-\omega_{m}$. Not much amplification is observed in the scattered light power. In both the figures (\ref{d}) and (\ref{e}), we observe that at a particular $\Delta_{c}$, the plots for the scattered number of photons and corresponding mirror displacement are similar i.e. they are not effected by the strong or weak modulation. Also, the amplitude of the scattered number of photons inside the cavity is high for $\Delta_{c}=\omega_{m}$ than for $\Delta_{c}=-\omega_{m}$.

Now, we will describe the experimentally realizable parameters used in our calculations. A BEC consisting of $10^5$ $^{87}Rb$ atoms \citep{cam} coupled to an optical field of an ultra high finesse Fabry Perot cavity has the strength of the coherent coupling as $g_{0} = 2 \pi  \times 10.9$ MHz \citep{nag}, ($2 \pi \times 14.4 MHz$ \citep{mur}). The kinetic and potential energy of the atoms is about $\nu = 35$ kHz \citep{bre} ($\nu = 49$ kHz \citep{mur}). Also the coherent amplification or the damping of atomic motion is ignored as the condensate temperature $T_{c} <<  \hbar \gamma_{c} /k_{B}$. The detuning of the atom pump is $2 \pi \times 32$ GHz. The atom field interaction is decreased due to decrease in the energy of the cavity mode because of the loss of photons through the cavity mirrors. By using high quality factor cavities, such loss of photons can be minimized. The cavity field can have damping rate $\kappa = 2\pi \times 8.75$ kHz \citep{nag} ($2\pi \times 0.66$ MHz \citep{mur}). The mechanical frequency of themovable mirror in an ptomechanical system can be altered from $2\pi \times 100$ Hz \citep{gar}, $2\pi \times 10$ kHz \citep{hun}, to $2\pi \times 73.5$ MHz \citep{sch2}. The corresponding decay rate of the movable mirror can thus be varied from $2\pi \times 10^{-3}$ Hz \citep{gar}, $2\pi \times 3.22$ Hz \citep{hun}, to $2\pi \times 1.3$ kHz \citep{sch2}.  

\section{Conclusion}

In conclusion, we have analyzed how the superradiant light scattering from an elongated Bose-Einstein condensate in an optomechanical cavity in the quantum regime is altered in the presence of the dynamical Casimir effect by considering the classical and quantized motion of the movable mirror. The cavity frequency is sinusoidally modulated with time in both mirror motions. The scattered light spectrum shows Bragg oscillations, modified by the dynamical Casimir effect, superimposed on the usual superradiant spectrum. The mirror motion has a dispersive effect leading to higher amplitude of scattered photon for lower modulation amplitude at initial times. Also, the amplitude of scattered photon number is much higher for classical mirror motion than the quantized mirror motion. The quantized motion of the movable mirror leads to similar behaviour of the spectrum of scattered light for both weak and strong modulations. The dynamical Casimir effect is found to be a new handle to coherently control and manipulate the superradiant light scattering process. 

\section{Acknowledgements}

Sonam Mahajan acknowledges University of Delhi for the University Teaching Assistantship. A. Bhattacherjee and Neha Aggarwal acknowledge financial support from the Department of Science and Technology, New Delhi for financial assistance vide grant SR/S2/LOP-0034/2010. 

\section{Appendix A}

In this appendix, we derive the time-dependent Hamiltoninan for the system by considering the quantized motion of the movable mirror. As one knows that, the cavity frequency is given as \citep{dod1}

\begin{eqnarray}
\omega_{c}=\frac{C_{1}}{L},
\end{eqnarray}

where $C_{1}$ is constant and $L$ is the length of the cavity. Therefore, change in the cavity length modifies the cavity frequency. Hence, the time varying cavity frequency is given as 

\begin{eqnarray}
\omega_{c}(x,t)=\frac{C_{1}}{L-x(t)}.
\end{eqnarray}

Since the perturbation in the cavity length is very small as compared to its original length. Therefore the time varying cavity frequency becomes

\begin{eqnarray}\label{one}
\omega_{c}(x,t)=\omega_{c}\left(1+\frac{x(t)}{L}\right).
\end{eqnarray}

Now one can take the time-modulated perturbation in cavity length as 

\begin{eqnarray}
x(t)=x'\epsilon '\sin\left(\Omega t\right).
\end{eqnarray}

where $x'=(\Delta x)\left(b+b^{\dagger}\right)$, $x'$ is the quantized perturbation in cavity length, $b$ is the annhilation operator of movable mirror, $\left(b+b^{\dagger}\right)$ is the dimensionless position operator of the movable mirror, $\epsilon '$ is the modulation amplitude and $\Omega$ is the modulation frequency. Using the time-modulated perturbation of cavity length in equation \ref{one}, we get

\begin{eqnarray}
\omega_{c}(x,t)=\omega_{c}\left(1+\frac{\left(\Delta x\right)\left(b+b^{\dagger}\right)\epsilon '\sin\left(\Omega t \right)}{L}\right).
\end{eqnarray}

Now we take normalized modulation amplitude as $\epsilon=\left(\left(\Delta x\right)\epsilon '\right)/L$, therefore the time varying cavity frequency becomes

\begin{eqnarray}\label{two}
\omega_{c}(t)=\omega_{c}\left(1+\left(b+b^{\dagger}\right)\epsilon\sin\left(\Omega t \right)\right).
\end{eqnarray}

The effective frequency also changes in this case as it is dependent on time varying cavity frequency (see eqn \ref{4a}). Therefore, modified effective frequency for the system with the quantized mirror motion becomes

\begin{eqnarray}\label{three}
\chi'(t)=\frac{\left(\epsilon'\Omega\cos\left(\Omega t\right)\right)}{4}=\chi(t)\left(b+b^{\dagger}\right).
\end{eqnarray}

The coupling constant for the photon-exciton coupling is given as \citep{dod1}

\begin{eqnarray}
g_{0}=\frac{C_{2}}{\sqrt{L}},
\end{eqnarray}

where, $C_{2}$ is a constant. Now the time varying coupling parameter becomes

\begin{eqnarray}
g_{0}(x,t)=\frac{C_{2}}{\sqrt{L-x(t)}}.
\end{eqnarray}

Under the small cavity length perturbation, the time varying coupling parameter becomes

\begin{eqnarray}
g_{0}(x,t)=\frac{C_{2}}{\sqrt{L}}\left(1+\frac{x(t)}{2L}\right).
\end{eqnarray}

Again, using the small quantized perturbation of the cavity length, we get

\begin{eqnarray}
g_{0}(x,t)=g_{0}\left(1+\frac{\left(\Delta x\right)\left(b+b^{\dagger}\right)\epsilon '\sin\left(\Omega t \right)}{2L}\right).
\end{eqnarray}

Therefore, the time varying coupling parameter becomes

\begin{eqnarray}\label{four}
g_{0}(t)=g_{0}\left(1+\frac{\left(b+b^{\dagger}\right)\epsilon\sin\left(\Omega t \right)}{2}\right).
\end{eqnarray}

Hence, the Hamitonian for the system with the quantized mirror motion under rotating-wave and dipole approximation can be written as

\begin{eqnarray}\label{five}\nonumber
H_{II} &=& \hbar\omega_{c}(t)a^{\dagger}a+\int dz\left\lbrace \psi _{g}^{\dagger}(z)\left(-\frac{\hbar ^{2}}{2m}\frac{\partial ^{2}}{\partial z^{2}}\right)\psi _{g}(z)+ \psi _{e}^{\dagger}(z)\left(-\frac{\hbar ^{2}}{2m}\frac{\partial ^{2}}{\partial z^{2}} + \hbar \omega _{a}\right)\psi _{e}(z)\right\rbrace\nonumber \\ &+& i\hbar g_{0}(t)a^{\dagger}\left\lbrace \int dz \psi _{g}^{\dagger}(z)\left(e^{-ikz}+e^{ikz}\right) \psi _{e}(z)\right\rbrace -i\hbar g_{0}(t)a\left\lbrace \int dz \psi _{e}^{\dagger}(z)\left(e^{ikz}+e^{-ikz}\right) \psi _{g}(z)\right\rbrace\nonumber \\ &-& i\hbar \eta'\left(a-a^{\dagger}\right)+i\hbar\chi '(t)\left(a^{\dagger 2}e^{-2i\omega_{p}t}-a^{2}e^{2i\omega_{p}t}\right)\left(b+b^{\dagger}\right)+\hbar\omega_{m}b^{\dagger}b.
\end{eqnarray} 

In the above Hamiltonian, the free energy of the movable mirror is represented by $\hbar\omega_{m}b^{\dagger}b$ where $\omega_{m}$ is the frequency. Now substituting the time varying cavity frequency (eqn. \ref{two}), time varying effective frequency (eqn. \ref{three}) and time varying coupling parameter (eqn. \ref{four}) in the above Hamiltonian (eqn. \ref{five}), the Hamiltonian becomes

\begin{eqnarray}\label{six}\nonumber
H_{II} &=& \hbar\omega_{c}a^{\dagger}a+\int dz\left\lbrace \psi _{g}^{\dagger}(z)\left(-\frac{\hbar ^{2}}{2m}\frac{\partial ^{2}}{\partial z^{2}}\right)\psi _{g}(z)+ \psi _{e}^{\dagger}(z)\left(-\frac{\hbar ^{2}}{2m}\frac{\partial ^{2}}{\partial z^{2}} + \hbar \omega _{a}\right)\psi _{e}(z)\right\rbrace\nonumber \\ &+& i\hbar g_{0}a^{\dagger}\left\lbrace \int dz \psi _{g}^{\dagger}(z)\left(e^{-ikz}+e^{ikz}\right) \psi _{e}(z)\right\rbrace -i\hbar g_{0}a\left\lbrace \int dz \psi _{e}^{\dagger}(z)\left(e^{ikz}+e^{-ikz}\right) \psi _{g}(z)\right\rbrace\nonumber \\ &-& i\hbar \eta'\left(a-a^{\dagger}\right)+i\hbar\chi(t)\left(a^{\dagger 2}e^{-2i\omega_{p}t}-a^{2}e^{2i\omega_{p}t}\right)\left(b+b^{\dagger}\right)+\hbar\omega_{m}b^{\dagger}b\nonumber \\ &+& \hbar k_{0}(t)a^{\dagger}a\left(b+b^{\dagger}\right)+i\hbar G(t)a^{\dagger}\left(b+b^{\dagger}\right)\left\lbrace \int dz \psi _{g}^{\dagger}(z)\left(e^{-ikz}+e^{ikz}\right) \psi _{e}(z)\right\rbrace\nonumber \\ &-& i\hbar G(t)a\left(b+b^{\dagger}\right)\left\lbrace \int dz \psi _{e}^{\dagger}(z)\left(e^{ikz}+e^{-ikz}\right) \psi _{g}(z)\right\rbrace .
\end{eqnarray} 

where $k_{0}(t)=\omega_{c}\epsilon\sin\left(\Omega t\right)$ and $G(t)=g_{0}\epsilon\sin\left(\Omega t\right)/2$ are the time-dependent coupling parameters.

\end{document}